\documentclass[prl,twocolumn,showpacs,floatfix,amsmath,amssymb,superscriptaddress]{revtex4-1}
\usepackage{amsfonts}
\usepackage{stmaryrd}
\usepackage{bbm}
\usepackage{mathrsfs}
\usepackage{tipa}
\usepackage{amssymb}
\usepackage{txfonts}
\usepackage{graphicx}
\usepackage{dcolumn}
\usepackage{epstopdf}
\usepackage[colorlinks,linkcolor=blue,urlcolor=blue,citecolor=blue]{hyperref}
\usepackage{multirow}
\usepackage{subfigure}
\usepackage{url}

\begin{document}
\newcommand*{\cm}{cm$^{-1}$\,}
\newcommand*{\Tc}{$T_v$\,}

\title{Dramatic change of photoexcited quasiparticle relaxation dynamics across Yb valence state transition in YbInCu$_4$}

\author{M. Y. Zhang}
\affiliation{International Center for Quantum Materials, School of Physics, Peking University, Beijing 100871, China}

\author{R. Y. Chen}
\affiliation{International Center for Quantum Materials, School of Physics, Peking University, Beijing 100871, China}

\author{T. Dong}
\affiliation{International Center for Quantum Materials, School of Physics, Peking University, Beijing 100871, China}

\author{N. L. Wang}
\email{nlwang@pku.edu.cn}
\affiliation{International Center for Quantum Materials, School of Physics, Peking University, Beijing 100871, China}
\affiliation{Collaborative Innovation Center of Quantum Matter, Beijing, China}

\begin{abstract}
    YbInCu$_4$ undergoes a first order structural phase transition near $T_v$=40 K associated with an abrupt change of Yb valence state. We perform ultrafast pump-probe measurement on YbInCu$_4$ and find that the expected heavy fermion properties arising from the \emph{c-f} hybridization exist only in a limited temperature range above $T_v$. Below $T_v$, the compound behaves like a normal metal though a prominent hybridization energy gap is still present in infrared measurement. We elaborate that those seemingly controversial phenomena could be well explained by assuming that the Fermi level suddenly shifts up and becomes far away from the flat \emph{f}-electron band as well as the indirect hybridization energy gap in the mixed valence state below $T_v$.

\end{abstract}

\maketitle

Heavy fermion or mixed-valent systems containing rare earth and actinide elements with partially filled 4$f$- or 5$f$-electron shells are among the most fascinating materials in condensed matter physics. The strong correlation of $f$-electrons and their coupling to the conduction electrons lead to a wide range of exotic phenomena. Depending on temperature the $f$-electrons show both itinerant and localized behaviors. At high temperatures, e.g. well above the Kondo temperature $T_K$, $f$-electrons are essentially localized and form local magnetic moments. With decreasing \emph{T}, the Kondo interaction becomes progressively stronger and it causes the formation of a flat \emph{f}-electron band near the Fermi level which hybridizes with bands from conduction electrons (\emph{c-f} hybridization). At $T\ll T_K$, the local moments are well-screened and $f$-electrons become itinerant. In the mean time a hybridization energy gap develops.

\begin{figure}
\includegraphics[clip,scale=0.63]{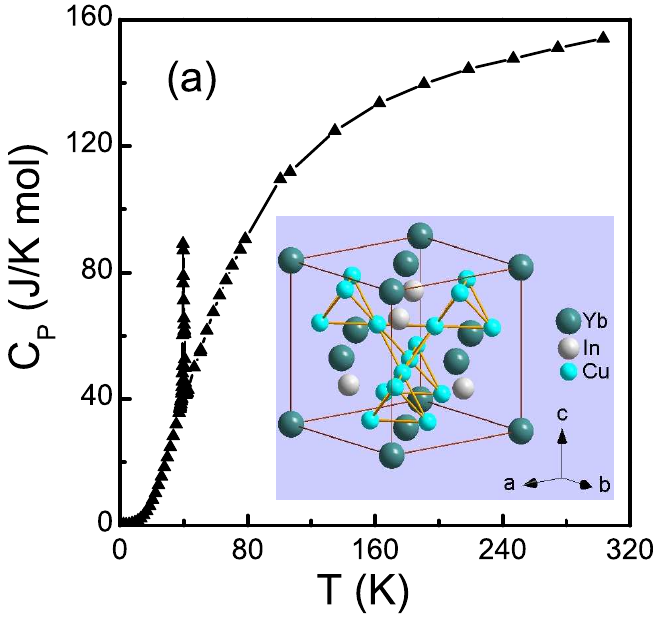}
\includegraphics[clip,scale=0.62]{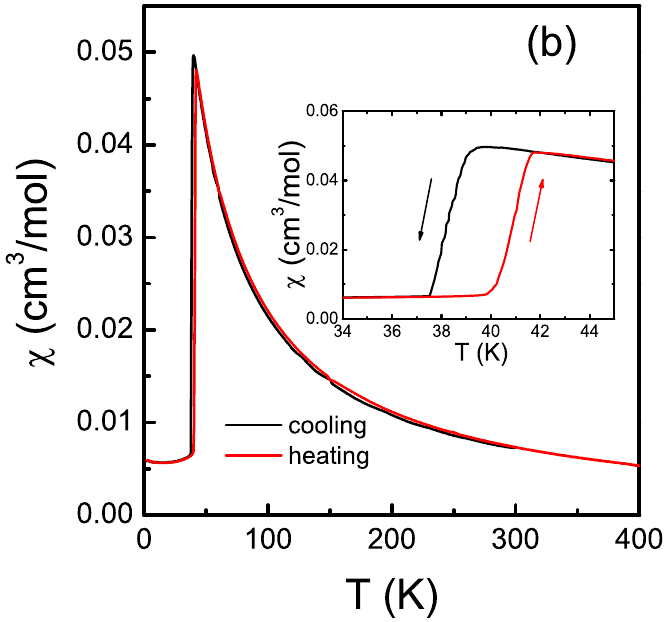}
\caption{(a) Temperature dependent specific heat of YbInCu$_4$. A sharp peak is seen near 40 K. Inset shows the crystal structure of  YbInCu$_4$.  (b) Magnetic susceptibility versus temperature for  YbInCu$_4$. The cooling and heating has a hysteresis due to the first order phase transition. When cooling from high temperature
the susceptibility shows a sharp drop near 39 K, and reaches a constant value below 37.5 K.}\label{Fig:MH}
\end{figure}

Usually, the temperature-induced crossover between localized and itinerant regimes is continuous, as exemplified by a broad maximum in the T-dependence of susceptibility and a crossover towards coherent transport with significantly reduced resistivity. However, a few intermetallic compounds exhibit discontinuous phase transitions. A notable example is YbInCu$_4$ which undergoes a first order structural phase transition ($T_v$) near 40 K \cite{Felner1986First,Kojima1990Neutron}. The compound has a face-centered cubic structure as shown in the inset of Fig. \ref{Fig:MH} (a). It behaves much differently from other YbXCu$_4$ series with X=Ag, Au, Cd,, Mg, Tl, and Zn \cite{Antonov,Sarrao1999Physical}. The transition is accompanied with an abrupt change of Yb valence. In the high temperature (HT) phase above $T_v$, Yb is approximately trivalent which has local magnetic moments and displays a Curie-Weiss susceptibility. While in the low temperature (LT) phase below $T_v$, the Yb valence is reduced to about 2.85 and the moment vanishes suddenly leading to an enhanced Pauli paramagnetic susceptibility\cite{Dallera2002New}. Abrupt changes were also observed in other physical quantities such as electrical resistivity, magnetization, specific heat capacity and nuclear spin-lattice relaxation rate 1/T$_1$ \cite{Cornelius1997Experimental,Kishimoto2003Mixed}. The valence transition is similar to the $\alpha-\beta$ transition of Ce\cite{Allen}, but it cannot be described by the Kondo volume-collapse model since the volume change is too small (about 0.5\%).

It is widely accepted that the disappearance of the local moment below the transition in YbInCu$_4$ is due to the Kondo screening effect \cite{Hancock2004Kondo, Sarrao1996Evolution, Jarrige, Chen2016Infrared}. The Kondo temperature $T_K$, calculated from susceptibility $\chi(T)$ under an impurity limit, increase abruptly from 25 K in the high-temperature phase above $T_v$ to over 400 K in the low-temperature phase below $T_v$ \cite{Sarrao1998Thermodynamics}. Therefore, YbInCu$_4$ offers an ideal system to study the change of charge excitation properties with different Kondo temperatures or hybridization strengths. Here we report an ultrafast optical pump optical probe measurement on YbInCu$_4$. A rapid increase in both relaxation time and amplitude of the ultrafast photoinduced reflectivity with temperature decreasing was observed above $T_v$, which are the characteristic features of typical heavy fermion materials. However, the relaxation time suddenly changed back to a very small value and displayed almost no temperature dependence once below the structural and valence phase transition. It behaves as a typical response of conventional metal as if the hybridization effect suddenly collapses. We further performed optical spectroscopy measurement and confirmed the presence of a large hybridization energy gap arising from\emph{ c-f } hybridization, similar to earlier reports in literature. We address those seemingly controversial phenomena by proposing that, in the mixed valence state below $T_v$, the Fermi level suddenly shifts up and becomes far away from the hybridization energy gap as well as the flat \emph{f}-electron band.

 The high-quality single crystals of YbInCu$_4$ were grown by InCu flux method\cite{Felner1986First,Kojima1990Neutron} and characterized by specific heat and magnetic susceptibility measurements. Abrupt changes of physical properties were observed clearly near 40 K \cite{Felner1986First}. The specific heat measured in a Quantum Design Physical Properties Measurement System (PPMS) indicates a sharp peak at 40 K as shown in Fig. \ref{Fig:MH} (a), yielding evidence for a first order phase transition. The magnetic susceptibility $\chi$ as a function of temperature measured also in the PPMS system is displayed in Fig. \ref{Fig:MH} (b). It shows a Curie-Weiss like behavior at high temperature above the phase transition but drops sharply to a nearly temperature-independent value. A hysteresis in magnetization was clearly seen when cooling down the sample and then warming it up, providing further evidence for the first order phase transition. In the cooling process, the onset of the transition is close to 39 K, while the transition completes at the temperature of 37.5 K, as seen clearly from the inset of Fig. \ref{Fig:MH} (b). The observations are similar to the published data in the literature. Those characterizations demonstrate the good quality of our single crystal samples.

The transient reflectivity of the compound was measured with a standard pump-probe set up. A Ti:sapphire oscillator was utilized as the source of both pump and probe beams and produces 800-nm laser pulses with 100 fs duration at a 80 MHz repetition rate. The pump beam was modulated with a frequency of 1 MHz and polarized perpendicular to the probe beam. We kept the pump fluence at about 0.8 $\mu$J/cm$^2$ ,10 times stronger than the probe pulses to minimize the overall heating of the sample.

\begin{figure}[htbp]
  \centering
 \includegraphics[scale=0.5,width=7.5cm]{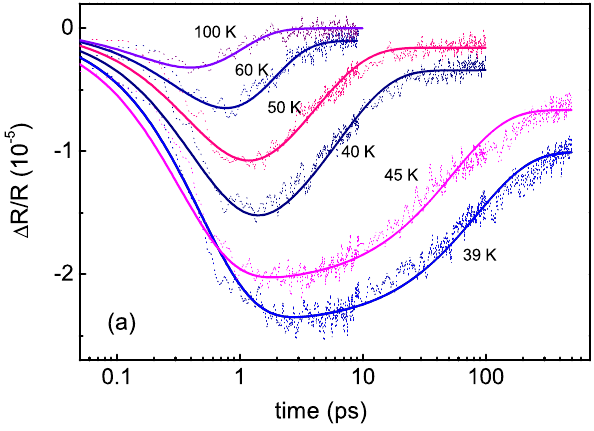}
  \includegraphics[scale=0.5,width=7.5cm]{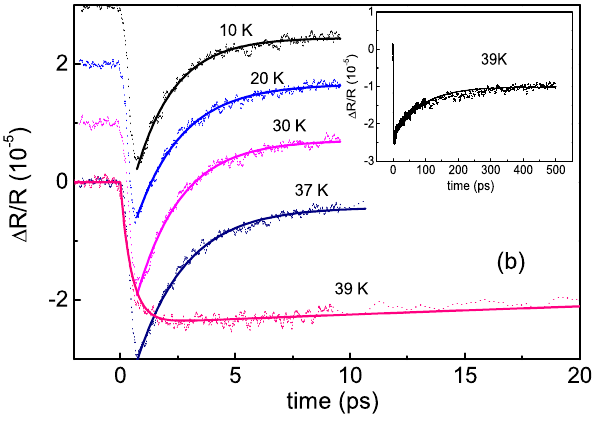}
  \caption{Photoinduced reflectivity data on  YbInCu$_4$ at (a) T$>T_v$, and (b) T$<T_v$. For comparison, the data just above the transition (at 39 K) is also shown in (b). The experimental data are presented by dots, while theoretical fits are shown as solid lines. The data on (b) have been vertically shifted for clarity. }\label{Fig:RT}
\end{figure}

Figure \ref{Fig:RT} presents the photoinduced (PI) reflectivity on a cleaved YbInCu$_4$ at a number of selective temperatures. The data were collected when the sample was cooled down from high temperature. A sudden change was found precisely at the phase transition temperature. We checked by changing the fluence of pump beam that the steady-state heating effects can be neglected in our measurement \cite{Demsar2006}. In Fig. \ref{Fig:RT} (a), the quasiparticle dynamics are strongly temperature dependent in the HT phase above $T_v$. The amplitude of the photoinduced reflectivity $\Delta R/R$ is substantially enhanced with temperature cooling and the decay time gets much longer at the same time. It is well known that the slowing down of the relaxation at low temperatures is a generic feature of the heavy electron systems \cite{Demsar2003Quasiparticle}. As we shall elaborate below in detail, the observed behavior is indeed due to the enhanced Kondo screening effect. On the other hand, when the temperature drops below $T_v$ (e.g. 37 K), the relaxation time suddenly decreases from about hundred of picoseconds to around 2 ps, as shown in Fig. \ref{Fig:RT} (b). With further decreasing temperature, the transient reflectivity displayed essentially no temperature dependence. It behaves as a typical response of conventional metal as if the hybridization effect disappeared below $T_v$. The measurement reveals significantly different electronic states above and below $T_v$ in YbInCu$_4$.

\begin{figure}[htbp]
  \centering
  \includegraphics[scale=0.5,width=4.3cm]{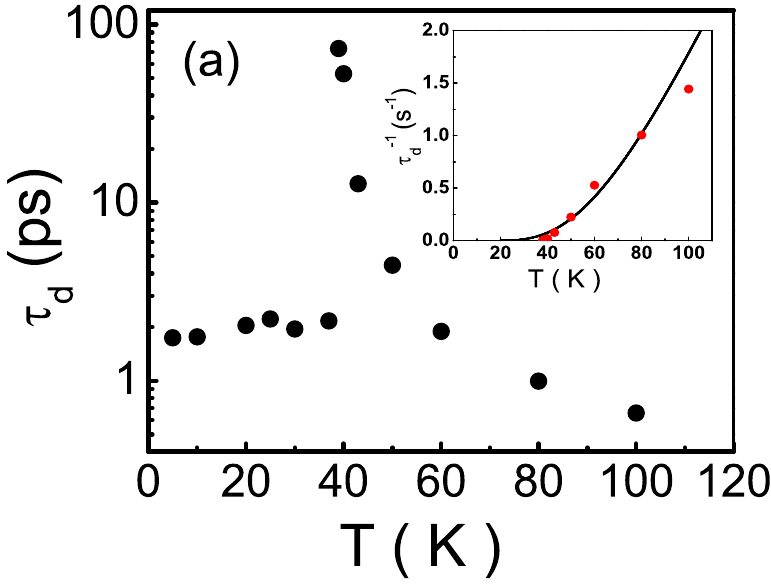}
  \includegraphics[scale=0.5,width=4.1cm]{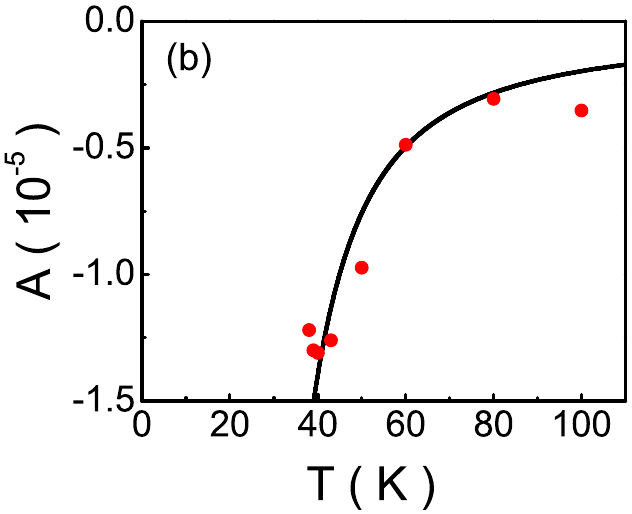}
    \includegraphics[scale=0.5,width=4.2cm]{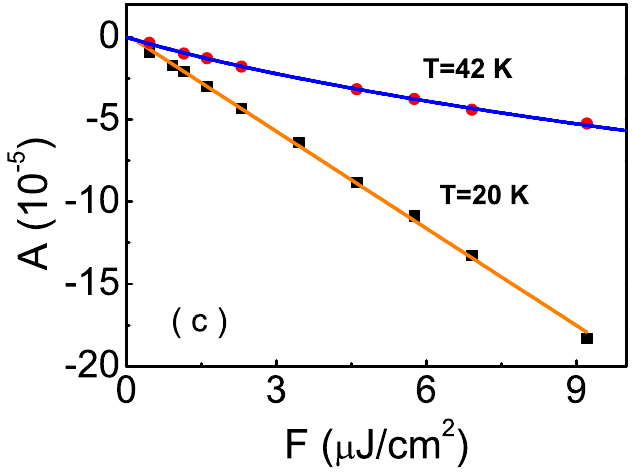}
   \includegraphics[scale=0.5,width=4.2cm]{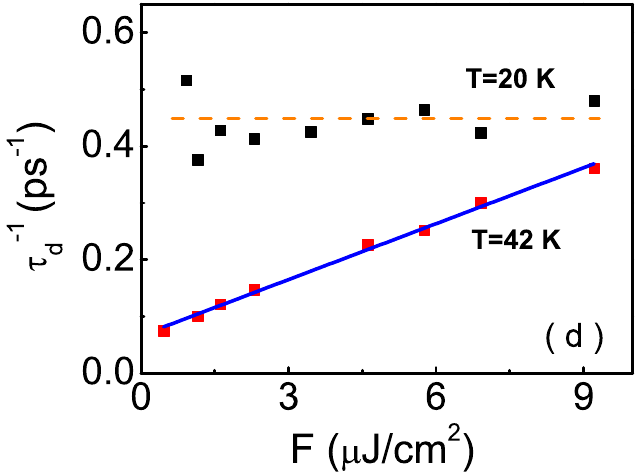}
  \caption{ (a) The temperature-dependent decay time $\tau^{-1}$. Inset: the inverse of decay time above the valence transition temperature. (b) the transient amplitude $A$ above the valence transition temperature. The red circle dots are the experimental results and the black lines are the fit to the RT model. (c) and (d): photoexcitation fluence dependence of the transient amplitude $A$ and relaxation time $\tau^{-1}$ taken at the temperature T=20 K (LT phase) and T=42 K (HT phase) together with fit by the two temperature model and the Rothwarf-Taylor bottleneck model respectively. }\label{Fig:AR}
\end{figure}

We now make a quantitative analysis of the relaxation of the photo-excited quasiparticles. As shown in Fig. \ref{Fig:RT}, $\Delta R/R$ initially decreases to a maximum then increases back to the equilibrium state, the latter of which can be well described by a single exponential decay $\Delta R/R$ = $A$ exp$(-t/\tau_d)$+$C$, where $A$ stands for the amplitude of the photoinduced reflectivity, $\tau_d$ is the relaxation time of the decay and $C$ is the long lived bolometric signal corresponding to a heat diffusion process that usually takes several nanoseconds. In order to account for the long rise time $\tau_r$ in HT phase, we fit the photoinduced reflectivity $\Delta R/R$ in Fig. \ref{Fig:RT} (a) by $\Delta R/R$ = $\mu(t)$[1- exp$(-t/\tau_r)$][$A$ exp$(-t/\tau_d)$+$C$]. Here $\mu(t)$ is the unit step function and 1- exp$(-t/\tau_r)$ is used to describe the rising edge of the photoinduced transient \cite{Demsar2006}. While below $T_v$ we only fit the relaxation decay part since the rising edge is sharp and $\tau_r$ is constant, as displayed in Fig. \ref{Fig:RT} (b). The temperature dependence of the obtained decay time is plotted in Fig. \ref{Fig:AR} (a). It is obvious to see that the compound shows totally different behaviors above and below the first-order valence transition temperature. In the HT phase, the relaxation time continuously increases from sub-picosecond at 100 K to almost 100 ps at 39 K, but becomes almost temperature-independent at LT phase with $\tau_d \sim$2 ps.

The slowing down of the decay time and enhancement of amplitude were widely observed in superconductors and density wave systems near transition temperatures due to the formation of energy gaps in the excitation spectra and were well explained by the so-called phonon bottleneck effect of Rothwarf-Taylor (RT) model \cite{Rothwarf1967}. In the heavy fermion systems, the Kondo interaction would lead to the formation of a hybridization energy gap below the coherent temperature, the phonon bottleneck effect was also widely used to explain the substantial increase of the relaxation time and enhancement of amplitude of photoexcited quasiparticle decay process \cite{Hewson1997The,Logan2005Dynamics,Demsar2006}. Nevertheless, different from the phase transitions in superconductors or charge density wave systems, the  hybridization of the conduction band and \emph{f} electrons in heavy fermion systems becomes progressively stronger at low temperature and is a crossover phenomenon rather than a second order phase transition. In this circumstance, the T dependence of the amplitude $A(T)$ and relaxation time $\tau_d(T)$ of the PI reflectivity can be described by the RT model by assuming a constant value of energy gap \cite{Demsar2006}. Within this mode \cite{Demsar2006,Kabanov2005}, $A(T)$ is given by
\begin{equation}\label{Eq:A}
  n_T(T)\simeq T^{1/2}exp(-E{g}/2T)\propto A^{-1}-1,
\end{equation}
and the relaxation time $\tau_d (T)$  follows that
\begin{equation}\label{Eq:tau}
  \tau^{-1}(T)=\Gamma[\delta(\epsilon n_T+1)^{-1}+2n_T],
\end{equation}
where $n_T$ represents the density of thermally excited quasiparticles which is related to the energy scale of the hybridization gap $E_g$. $\Gamma$, $\delta$, $\epsilon$ are all fitting parameters that are independent of temperatures. We find that $\tau_d$ (inset of Fig. \ref{Fig:AR} (a)) and $A$ (Fig. \ref{Fig:AR} (b)) follow Eq.\ref{Eq:tau} and Eq.\ref{Eq:A} in the HT phase quantitatively and the gap energy is determined to be $E_g$$\approx$300 K.

To further clarify the difference between electronic states above and below the structural and valence phase transition, we also performed excitation intensity dependent studies of the relaxation dynamics with an excitation fluence($F$) ranging from 0.4 to 9 $\mu$J/cm$^2$  as shown in Fig. \ref{Fig:AR} (c) and (d). In the HT phase, the transient amplitude is fitted well by $A\propto \sqrt{1+c F}$-1, a formula derived from Rothwarf-Taylor model \cite{Demsar2006}, with \emph{c} =0.4. $\tau^{-1}$ shows a linear $F$-dependence at T=42K, also matching the RT model's deduction. Similar behaviors of $A(F)$ and $\tau^{-1}$($F$) have been observed in many other heavy fermion systems like YbAgCu$_4$, SmB$_6$ as described by \cite{Demsar2006} . On the other hand, in the LT phase (T=20K), $A$ exhibits linear $F$-dependence while $\tau^{-1}$ is basically independent of $F$. Such fluence dependent properties for $A$ and $\tau^{-1}$ at low temperatures are common features of normal metal like Ag and Au which can be well explained by considering both the electron-electron and electron-phonon thermalizations (two temperature model) \cite{Allen1987Theory,Ahn2003Ultrafast,Demsar2003Quasiparticle}.

\begin{figure}[htbp]
 \centering
  \includegraphics[scale=0.5,width=8.8cm]{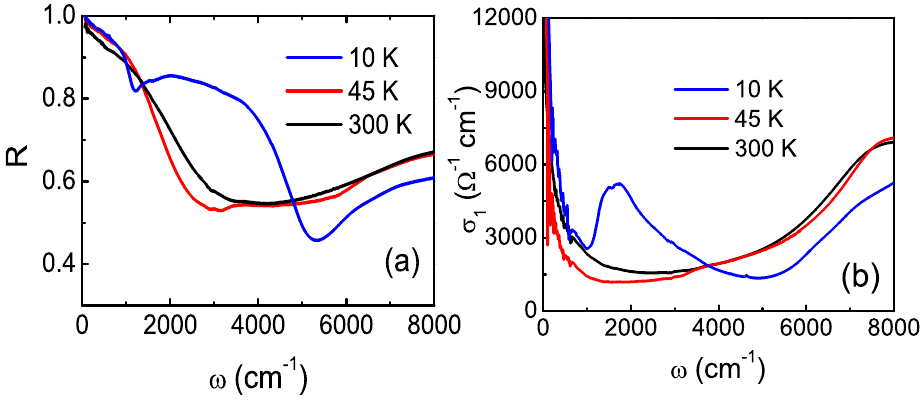}
 \caption{Optical reflectance (a) and conductivity (b) spectra for YbInCu$_4$ above and below $T_v$. }\label{Fig:IR}
\end{figure}

The above study on the relaxation dynamics of photoexcited quasiparticles seems to suggest that the \emph{c-f} hybridization exists only in a limited temperature range above the structural phase transition but collapses below the Yb valence transition. However, this conclusion is in contrast to the experimental fact that the YbInCu$_4$ is in the mixed valence state below $T_v$ with much higher Kondo temperature where even stronger hybridization effect is expected. To demonstrate whether the hybridization energy gap really exists below $T_v$ in the present compound, we further performed optical reflectance measurement over broad energy scale from about 50 \cm to 45000 \cm in Bruker 80 v and grating-type spectrometers, and extracted optical conductivity through Kramers-Kronig transformation of reflectance data. The reflectance and conductivity spectra at several selected temperatures were presented in Fig.\ref{Fig:IR} (a) and (b). A dramatic change over broad energy scale was observed in conductivity spectra when the sample is cooled through the phase transition temperature. A pronounced peak near 1800 \cm was present at low T phase. The observations are similar to the published data in the literature \cite{Hancock2004Kondo, Garner2000Optical,Okamura2007a}.

One has to reconcile the observation in ultrafast quasiparticle relaxation dynamics with that in the infrared spectroscopy measurement. A naive picture is that the heavy fermion properties arising from the \emph{c-f} hybridization exists only in a limited temperature range above $T_v$. Below $T_v$,  the \emph{c-f} hybridization effect really disappears. Then, the sudden change of the optical conductivity over broad energy scale would be caused by the band structure reconstruction after the first order structural phase transition. The pronounced mid-infrared peak would reflect a new interband transition after the band reconstruction and is not relevant to the \emph{c-f} hybridization. However, this possibility is unlikely since many other measurements actually demonstrate that the compound is in mixed valence state where stronger hybridization effect is present below $T_v$.

\begin{figure}[htbp]
 \centering
\includegraphics[scale=0.5,width=8cm]{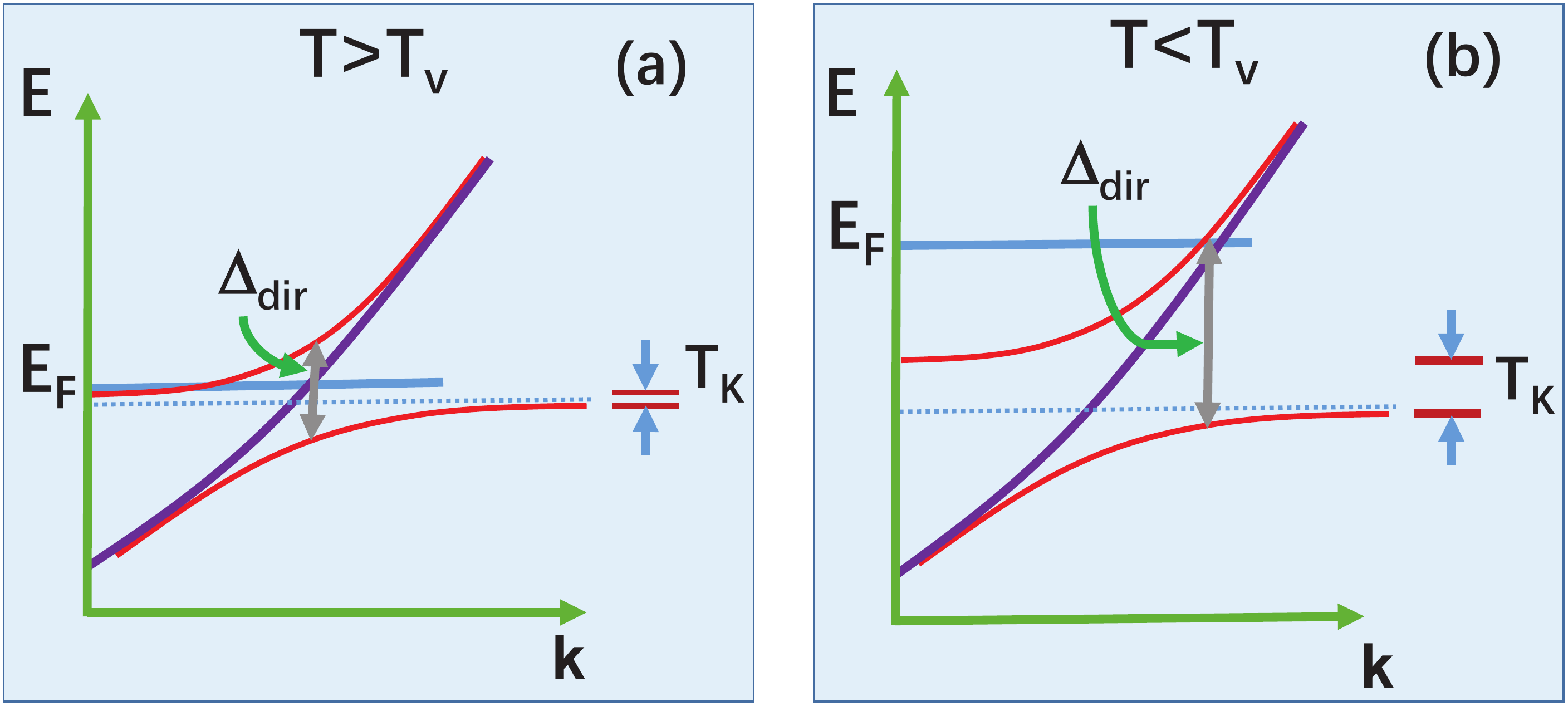}
 \caption{A schematic picture for YbInCu$_4$ on the basis of \emph{c-f} hybridization effect. Above $T_v$, YbInCu$_4$ is in the heavy fermion state with a small $T_K$ and the Fermi level is very close to the flat band. Below $T_v$, YbInCu$_4$ is in the mixed valence state with an enhanced $T_K$.  The Fermi energy shifts up due to the increased filling of Yb \emph{f-}electrons and is away from the flat \emph{f}-electron band. }\label{Fig:Schem}
\end{figure}

A more plausible picture is that the pronounced mid-infrared peak in the optical conductivity spectrum below $T_v$ indeed reflects the strong \emph{c-f} hybridization effect in the mixed valence state and is an indication of hybridization energy gap as suggested in earlier studies. Then we need to understand why the ultrafast quasiparticle relaxation dynamics resembles very much to that in a normal metal? We find that the experimental observations could be well explained by assuming that, in the mix valence state below $T_v$, the fermi level suddenly shifts up and becomes far away from the flat \emph{f}-electron band arising from the Kondo interaction or \emph{c-f} hybridization based on the periodic Anderson lattice model (PAM) \cite{Hancock2004Kondo, Chen2016Infrared}.

As shown in Fig. \ref{Fig:Schem}, the Kondo interaction or \emph{c-f} hybridization would lead to the formation of a hybridization energy gap in \textbf{k}-space with a direct gap of $\Delta_{dir}=2V$, where V is the hybridization strength in PAM.
Meanwhile, an indirect gap would appear in the density of states. It has much smaller energy scale with the order of $T_K$. The infrared spectroscopy detects the direct hybridization energy gap. In the heavy fermion state, the hybridization is relatively weak (with smaller $T_K$), the Fermi level is very close to the flat band, giving rise to the heavy effective mass. This is the situation for $T>T_v$ in the present case as illustrated in Fig. \ref{Fig:Schem} (a). Since the measurement temperature is higher than $T_K$, the hybridization energy gap feature is not visible in infrared measurement. We find that the ultrafast photoexcited quasiparticle relaxation dynamics is more sensitive to the hybridization effect. Upon a sudden change of Yb valence state below $T_v$, as shown in Fig. \ref{Fig:Schem} (b), the Kondo temperature $T_K$ abruptly increases, resulting in an enhanced indirect hybridization energy gap. Moreover, the Fermi energy shifts up due to the increased filling of Yb \emph{f-}electrons. Therefore, the Fermi level is away from the flat \emph{f}-electron band as well as the indirect energy gap. As a result, the charge carriers near the Fermi level have much smaller effective mass and behave more like normal conduction electrons. The photoexcited quasiparticles would decay via electron-electron and electron-phonon thermalizations without being impeded by a gap. Nevertheless, a larger direct hybridization gap $\Delta_{dir}$ could be probed by infrared spectroscopy.

In conclusion, we performed ultrafast pump probe measurements on YbInCu$_4$ single crystals and found that the heavy fermion characters exist only in a limited temperature range above the structural and valence phase transition. A rapid increase in both relaxation time and the amplitude with temperature decreasing was observed in the ultrafast photoinduced reflectivity dynamics above T$_v$. However, the relaxation time suddenly changed back to a very small value and displayed no temperature dependence once below $T_v$. The compound behaves like a normal metal though a prominent hybridization energy gap is still present in optical conductivity. We addressed those seemingly controversial phenomena and elaborated that the experimental observations could be well explained by assuming that the Fermi level suddenly shifts up and becomes far away from the flat \emph{f}-electron band as well as the indirect hybridization energy gap below $T_v$.

\begin{center}
\small{\textbf{ACKNOWLEDGMENTS}}
\end{center}

We acknowledge useful discussions with Yifeng Yang. This work was supported by the National
Science Foundation of China (No. 11327806), and the National Key Research and Development Program of China (No.2016YFA0300902).


\end{document}